# Lévy Diffusion and Classes of Universal Parametric Correlations

Dimitri Kusnezov[a, *] and Caio H. Lewenkopf[b, †]

[a] *Center for Theoretical Physics, Sloane Physics Laboratory, Yale University, New Haven, CT 06520-8120*
[b] *Department of Physics, FM-15, University of Washington, Seattle, WA 98195*

January 26, 1995

A general formulation of translationally invariant, parametrically correlated random matrix ensembles, is used to classify universality in correlation functions. Surprisingly, the range of possible physical systems is bounded, and can be labeled by a parameter $\alpha \in (0, 2]$, in a manner analogous to Lévy diffusion. Universality is obtained after scaling by the (anomalous) diffusion constant $D_\alpha$ (the usual scaling is divergent for $\alpha < 2$). For each $\alpha$, correlation functions are universal, and distinct. The previous results in the literature correspond to the limiting case of superdiffusion, $\alpha = 2$.

It is becoming increasingly possible to explore the physical properties of complex quantum systems with the tools offered by Random Matrix Theory (RMT). One of the main lessons learned from systematic studies of complex quantum many-body systems and simple quantum systems with a classical chaotic limit, is that they display the same kind of spectral universal fluctuations, depending only on the type of symmetry that they exhibit, e.g. orthogonal, unitary or symplectic [1]. Until very recently, however, very little was know about chaotic systems which depend on an external parameter $x$, and how the properties of such systems are correlated in that parameter. While the parameter is often viewed as an external quantity, such as an electric or magnetic field, it is also important in the study many-body systems with slow and fast modes, such as atomic nuclei. In that case, it corresponds to the slow variables in the adiabatic limit of the time-evolution of the many-body system. One of the remarkable properties which has emerged is that, under appropriate scaling, parametric-dependent correlation functions are universal as well. One of the first works addressing such question was a study of level curvature distributions [2]. A significant step towards the understanding of parametric correlations was recently achieved by Szafer, Simons and Altshuler [3,4] in a sequence of papers. They were able to compute analytically the density-density and level velocity correlators in a particular random matrix model realization, which is very successful in describing different chaotic and disordered systems. Correlation functions were found to fall upon universal curves, seemingly independent of the underlying properties of the studied systems. In this letter, we show that there are in fact a continuous number of universal curves for each type of observable, and that those known up to now are a just a limiting case. Furthermore, the quantity which is used to rescale the parameter to obtain universality, is generally divergent, requiring the introduction of a more general scaling.

The framework which we develop here is based on assuming that we have an ensemble of random matrix Hamiltonians $H(x)$, which depends on an external parameter $x$. At each value of $x$, the matrix has the same spectral properties, pertaining to the Gaussian Orthogonal Ensemble (GOE) (or equivalently the Gaussian Unitary (GUE) or Symplectic (GSE) Ensembles; we will consider here only the GOE case, as the others can be obtained in a straightforward manner). Furthermore, we ask that the system have translational invariance, in the sense that correlation functions at two different values of the external parameter $x, y$ depend only on their separation $|x - y|$. Hence, our starting point is the assumption that the covariance of $H$ is translationally invariant:

$$\overline{H_{ij}(x)H_{kl}(y)} = \frac{1}{2}F(x-y)(\delta_{ik}\delta_{jl} + \delta_{jk}\delta_{il}). \quad (1)$$

Here $F(x) = F(-x)$ defines the rate of decorrelation of the system as one varies the parameter. There are two independent formulations we use to realize such a $H - H$ correlation.

(I) <u>Stochastic Integral:</u> In this first approach, we consider the explicit construction of $H(x)$ as the most general linear combination of independent, uncorrelated GOE matrices:

$$H_{ij}(x) = \int dy\, f(x-y)V_{ij}(y). \quad (2)$$

Here $V_{ij}(y)$ is uncorrelated white noise at every $i, j, y$:

$$\overline{V_{ij}(y)} = 0, \quad \overline{V_{ij}(y)V_{kl}(y')} = \delta(y-y')(\delta_{ik}\delta_{jl} + \delta_{jk}\delta_{il}). \quad (3)$$

The stochastic integral (2) defines a theory such as (1) through the convolution

$$F(x-y) = \int dz\, f(x-z)f(y-z). \quad (4)$$

As $H(x)$ must be a member of the ensemble for every $x$, we see that $f(x)$ must be a real valued function.

(II) <u>Measure:</u> Alternately, we can derive a generic covariance (1) from the measure:

$$P(H)\mathcal{D}H \propto \prod_{i,j,x} dH_{ij}(x) \times \qquad (5)$$
$$\times \exp\left(-\frac{1}{2}\int\int dx\,dy\,\text{Tr}\,[H(x)K(x-y)H(y)]\right).$$

Since the measure is quadratic in $H$, the correlation function $F(x)$ is just the functional inverse of the operator $K$:

$$\int dz\,K(x-z)F(z-y) = \delta(x-y). \qquad (6)$$

The physical relevance of random matrix modelling is directly related with the identification of the scaling parameter with physical quantities. At the RMT level though, universality of correlation functions is typically obtained in the large $N$ limit by scaling the parameter $x$ with some power of $N$, where $N$ is the dimension of $H$. As a consequence, the understanding of the short distance behavior of $F(x-y)$ plays a central role. To explore this, we choose a natural, reasonable and fairly generic parametrization of the entire range of possible leading order short distance behaviors of $F(x-y)$ by choosing the function of the form:

$$F(x-y) = e^{-|x-y|^\alpha} \sim 1 - |x-y|^\alpha + \ldots \qquad (7)$$

Here $\alpha$ is an arbitrary real number. Surprisingly, we will see that the range of $\alpha$ cannot be arbitrary, but must be bounded. (As a result, parametric correlations are not generally related by a rescaling.) If we chose different realizations with the same small distance behavior, such as $F(x) = 1/(1+x^\alpha)$, $cos(x^{\alpha/2})$, and so forth, the conclusions we find are identical. In this sense the specific choice of $F$ is not important.

Consider first the range of $\alpha$ for which the function $K(x-y)$ defines a probability distribution. The measure, Eq. (5), is normalizable providing that for each pair $i,j$, the quantity

$$\int dx \int dy\,H_{ij}(x)H_{ji}(y)K(x-y) \geq 0, \qquad (8)$$

so that $K$ must be a positive definite operator. Since $F$ is the functional inverse of $K$, $F$ must also be positive definite. The properties of positive definite functions were systematically studied by Bochner [5]. Using the theorems of Bochner on positive definite functions [5], the only range of $\alpha$ in which $F(x)$ is positive definite is $0 < \alpha \leq 2$. Hence the measure is only defined for $\alpha \in (0,2]$.

The conclusion is the same from method (I): $F(x)$ can only be factorized into real functions $f$ in Eq. (4) when $F$ is a positive definite function. This is most easily seen in the special case when $f$ itself is symmetric, then

$$f(x-y) = \frac{1}{2\pi}\int dk\,\sqrt{F(k)}\cos(k(x-y)), \qquad (9)$$

where $F(k)$ is the fourier transform of $F(x-y)$. Again from Bochner we know that: the fourier transform $F(k)$ is a positive function, if and only if $F(x-y)$ is positive definite. Hence Eq. (9) is a well defined construction of $f(x-y)$ for $\alpha \in (0,2]$. As a consequence, it is interesting to note that one cannot choose a general covariance (1) and find a construction (2): $f$ can only be defined consistently when $\alpha \in (0,2]$. If $F$ is not positive definite, then $f$ is in general complex, and $H(x)$ is no longer a member of the chosen random matrix ensemble.

We have explicitly constructed parametric Hamiltonians $H(x)$ for $\alpha \in (0,2]$ by factorizing the positive definite function (7). In Fig. 1, we show a range of measured covariances of $H$, using matrices of dimension $N = 100$, for the values of $\alpha = \frac{1}{2}, 1, \frac{3}{2}, 2$ in Eq. (7). $H(x)$ was constructed over a range of $x$ for each case, and $F(x-y)$ was measured by averaging over $x$ and $y$, the quantity:

$$F(x-y) = \frac{1}{N(N-1)}\sum_{i<j} H_{ij}(x)H_{ij}(y). \qquad (10)$$

The results in Fig. 1 are representative points, the grid size being finer, and are in agreement with the expected value of $\alpha$. (For instance, a fit to the $\alpha = 3/2$ curve yields an exponent of $\alpha = 1.49(1)$.)

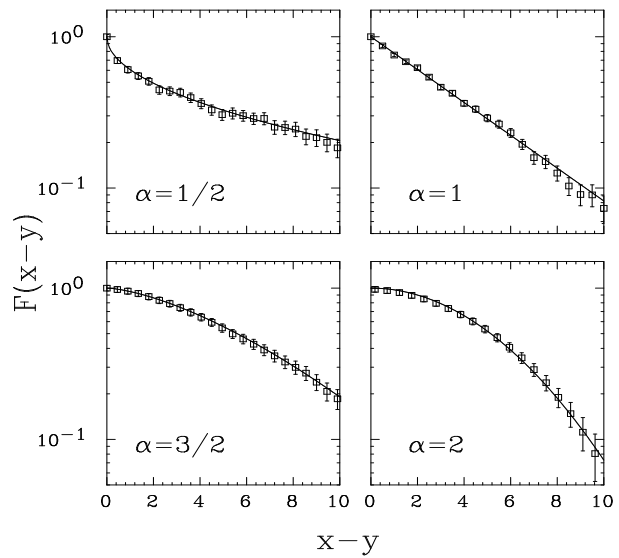

FIG. 1. Computed covariance $F(x-y) = \overline{H(x)H(y)}$ from simulations at $\alpha = 1/2, 1, 3/2, 2$ in (7). The solid line is the exact result and the boxes are the results of the simulation. The $x$ axis is rescaled in the figure in order to show the short range behavior. The points are shown are representative; the actual grid size is much smaller.

To explore the influence of $\alpha$ on the form of universal curves, we first consider the short distance behavior of the Hamiltonian as manifested in various observables. Using perturbation theory to explore the short distance properties, the energy and wavefunction overlaps, to second order, are

$$\delta E_n(x) = E_n(x') - E_n(x) = \qquad (11)$$
$$= \delta H_{nn} + \sum_{m \neq n} \frac{|\delta H_{mn}|^2}{E_n - E_m} + ...$$

$$|\langle \Psi_n(x')|\Psi_n(x)\rangle|^2 = 1 - \sum_{m \neq n} \frac{|\delta H_{mn}|^2}{(E_n - E_m)^2} + .... \qquad (12)$$

Using the ensemble average defined by the correlation (1), the short distance behavior (7), and that we have translational invariance, it follows from Dyson [6] that $\overline{(\delta E_n(x))^2} \sim \delta x^\alpha$, and $1 - \overline{|\langle \Psi(x')|\Psi(x)\rangle|^2} \sim \delta x^\alpha$. The case for $\alpha = 1$ corresponds to Dyson's Brownian motion model for random matrices [6]. One immediate observation is that not only do these quantities decorrelate differently for each $\alpha$, but $C(0)$, which is obtained from $C(x - x') = C(\delta x) = \overline{\frac{\partial E_i}{\partial x}(x)\frac{\partial E_i}{\partial x}(x')}$ for $\delta x \to 0$, diverges as $1/\delta x^{2-\alpha}$. By measuring $C(0)$ as a finite difference on a grid defined by the discretization of Eq. (2), the divergence will be in terms of the minimum grid spacing. Thus $C(0)$ is only finite for the special case $\alpha = 2$, and one cannot obtain universality by scaling the parameter $x$ by $\sqrt{C(0)}$. So, as $\delta x \to 0$:

$$C(0) \sim \begin{cases} \delta x^{\alpha-2} \to 0 & \alpha > 2 \\ \text{constant} & \alpha = 2 \\ (1/\delta x)^{2-\alpha} \to \infty & \alpha < 2 \end{cases} \qquad (13)$$

(Our simulations have confirmed this behavior for $\alpha \leq 2$.) It is clear that one must define a more general level velocity, and a new scaling. Because wavefunction correlations do not suffer from this behavior, they serve as a better measure of the anomalous character $\alpha$ of the universal behaviour. Eq. (13) provides another argument on the finite range of $\alpha$. If one had a realization of $\alpha > 2$, one would have both energy level correlations from (1) *and* a vanishing rms value of the energy level slope since $C(0) = 0$, which is not possible.

We would like to view the processes classified by $\alpha$ in terms of its similarities with Lévy or anomalous diffusion [7]. Anomalous diffusion of a particle in a disordered or chaotic environment generally leads to diffusion with a mean square deviation of the type

$$\langle Q^2(t)\rangle = Dt^\gamma, \qquad (14)$$

with $\gamma \neq 1$. When $\gamma = 1$, one recovers Brownian diffusion. In a similar manner, a natural extension of Dyson's Brownian motion model is then to interpret the behavior in terms of anomalous diffusion in the parameter (fictitious time):

$$\overline{\delta E^2} = D_\alpha \delta x^\alpha. \qquad (15)$$

Here $D_\alpha$ is defined as the anomalous diffusion constant. One would be tempted to object against this interpretation arguing that there is a fundamental difference between Lévy flights (14) and the parametric result (15):

The former is a *long-time* behavior, while the latter is related to *short-distance* in parameter space. This contradiction is only apparent, since the character of the considered gaussian ensembles magnifies enormously small perturbations which results in effectively scaling the $x$-parameter by some power of $N$. As a consequence, the "short distance" leading order behavior become the dominant feature as $N \to \infty$. In this sense, the large $N$ limit makes Eq. (15) analogous to Lévy diffusion. For $\alpha = 1$, we recover Brownian diffusion, for $\alpha > 1$ one has superdiffusion, while $\alpha < 1$ gives subdiffusion. Hence, the proper scaling of the coordinate must be related to the anomalous diffusion constant, which relates the average drift in energy to parametric distance:

$$\widetilde{x} \equiv [D_\alpha]^{1/\alpha} x, \qquad D_\alpha = \frac{\overline{\delta E^2}}{\delta x^\alpha}. \qquad (16)$$

Then $C(0)$ can be identified as the diffusion constant for the superdiffusive case $\alpha = 2$: $D_2 = C(0)$, so that $\widetilde{x} = \sqrt{C(0)} x$, recovering the universal scaling derived by Szafer, Simons and Altshuler [3].

In Fig. 2 we show the universal curves for the wavefunction overlap obtained for the processes of Fig. 1. When plotted as a function of the $\widetilde{x} = [D_\alpha]^{1/\alpha} x$, the decorrelations are universal, and distinct for each $\alpha$. Recall that scaling by $\sqrt{C(0)}$ does not result in universality since $C(0)$ is divergent for $\alpha < 2$. It is worth noting that the functional form of the wavefunction correlations are reproduced quite well by

$$\overline{|\langle\Psi(x')|\Psi(x)\rangle|^2} \approx \frac{1}{1 + c|\widetilde{x} - \widetilde{x}'|^\alpha} \qquad (17)$$

which is Lorentzian for $\alpha = 2$. (Numerical details, such as the nature of the constant $c$, will be discussed elsewhere).

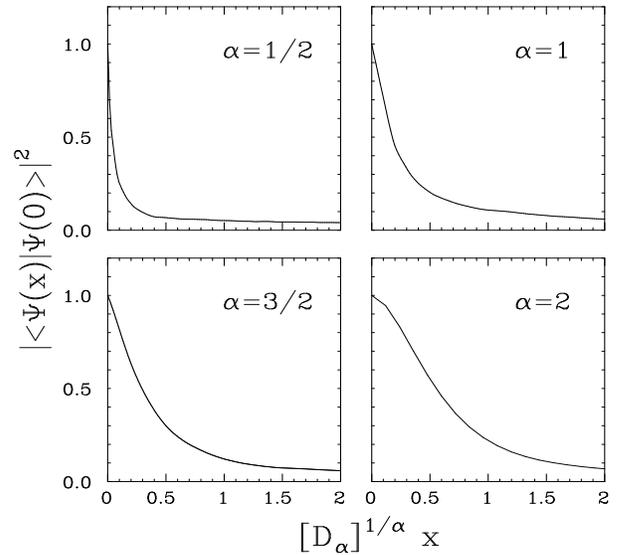

FIG. 2. Comparison of universal wavefunction overlaps for the different anomalous processes of Fig.1. Universality is obtained by rescaling with the diffusion constant: $\widetilde{x} = [D_\alpha]^{1/\alpha} x$.

We only point out here that the overlap $\overline{|\langle\Psi(x')|\Psi(x)\rangle|^2}$ provides a better measure of the value of $\alpha$ for a particular physical system, since it is not singular like $C(x)$. Hence, in this way, one can fit the functional form and measure $\alpha$.

There are several ways one can define a better measure of the level velocities. One is to define a generalized diffusion constant (an extension of the definition of $C(x)$ to include the stochastic measure for the anomalous process):

$$D(x-y) = \overline{\frac{\delta E(x)}{(\delta x)^{\alpha/2}}\frac{\delta E(y)}{(\delta x)^{\alpha/2}}}, \qquad D(0) = D_\alpha, \quad (18)$$

which for $\alpha = 2$, reduces to $D(x) = C(x)$. This function does not diverge on small scales. The form of $D(x)/D(0)$ for $\alpha = 1/2, 1, 3/2, 2$ is shown in Fig. 3. For $\alpha = 2$, one recovers the well known curve, but is otherwise different for different $\alpha$. Another is to introduce the fractional derivative of order $\alpha$ of the curve $E(x)$, denoted $\Delta_x^\alpha$, as a finite difference quotient. In the Grünwald-Liouville construction [8]

$$\Delta_x^\alpha E(x) = \lim_{\delta x \to 0}\frac{1}{\delta x^\alpha}\sum_{k=0}^\infty (-)^k \binom{\alpha}{k} E(x-k\delta x). \quad (19)$$

Then the natural definition for the scaling parameter is: $\widetilde{D}(x) = \overline{\Delta_x^{\alpha/2} E(x)\Delta_x^{\alpha/2} E(y)}$. Notice that in general $\widetilde{D}(0) \neq D_\alpha$.

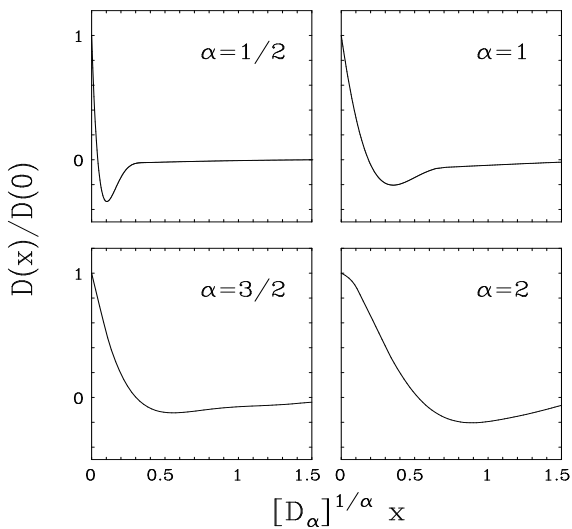

FIG. 3. The level velocity correlator $D(\widetilde{x})/D(0)$ versus $\widetilde{x}$ for $\alpha = 1/2, 1, 3/2, 2$. The results for $\alpha = 2$ are identical to those of Szafer, Simons and Altshuler Ref.[3]. (The $\alpha = 1/2$ case is numerically problematic since the slope of $F(x)$ at $x = 0$ diverges. For this reason the curve is more representative, having been smoothed with a spline.)

In conclusion, we have endeavored to formulate the most general class of correlated random matrix Hamiltonians which are translationally invariant. From this we shown that the range of possible short scale behaviours of the $H-H$ correlator is bounded, and can be assigned a value $\alpha \in (0,2]$. For every value of $\alpha$, there exists universal correlation functions. However, one of the surprising results is that these functions are distinct for each $\alpha$. They are not simply related by rescaling, due to the finite range of $\alpha$. The conventional universal scaling by $\sqrt{C(0)}$ was seen to diverge for all but the superdiffusive $\alpha = 2$. However, the introduction of scaling by the anomalous diffusion constant, $[D_\alpha]^{1/\alpha}$, was shown to be well defined for each $\alpha$, and result in universality for each $\alpha$. While we have constructed model Hamiltonians for different values of $\alpha$, it would be nice to find realizations in nature. One way to explore this is to measure wavefunction correlations, since these are non-singular, and one can extract the anomalous value $\alpha$ which governs the process. Finally, we note that when the parameter $x = x(t)$ is dynamic, the solution to the time-dependent Schrödinger equation for different observables depends on the short time behavior of the quantity $1 - F(t)$ [9], and we suspect that a similar classification exists for time dependent quantities such as occupation numbers.

This work was supported in part by the Department of Energy Grant DE-FG02-91ER40608. C. H. L. is supported by the National Science Foundation.

[1] See for example, O. Bohigas, in *Chaos and Quantum Physics*, Eds. M. Giannoni, A. Voros and J. Zinn-Justin, (North-Holland, New York, 1991).
[2] P. Gaspard, S. A. Rice, H. J. Mikeska and K. Nakamura, Phys. Rev. A **42**, 4015 (1990); J. Zakrzewsky and D. Delande, Phys. Rev. A **47**, 1650 (1993).
[3] A. Szafer and B. L. Altshuler, Phys. Rev. Lett. **70**, 587 (1993); B.D. Simons and B.L. Altshuler, Phys. Rev. Lett. **70**, 4063 (1993); Phys. Rev. B **48**, 5422 (1993).
[4] B.D. Simons, P.A. Lee and B.L. Altshuler, Phys. Rev. Lett. **70** 4122 (1993); ibid **72**, 64 (1994); Nucl. Phys. **B409**, 487 (1993).
[5] S. Bochner, *Lectures on Fourier Integrals*, (Princeton Univ. Press, Princeton, 1959).
[6] Y.Alhassid, D.Kusnezov and D.Mitchell, *Yale University, preprint* (1994); F.J. Dyson, J. Math. Phys. **3**, 1191 (1962).
[7] J.P.Bouchaud, A.Comtet, A.Georges and P.Le Doussal, Ann. Phys. (NY) **201**, 285 (1990); J.P.Bouchaud and A.Georges, Phys. Rep. **195**, 127 (1990); D. Kusnezov, Phys. Rev. Lett. **72**, 1990 (1994).
[8] B. Ross, *Fractional Calculus and its Applications*, (Springer-Verlag, New York, 1974); K. S. Miller and B. Ross, *An introduction to the fractional calculus and fractional differential equations*, (Wiley, New York, 1993).
[9] A.Bulgac, G. Do Dang and D.Kusnezov, *Yale University, preprint* (1995).